\documentclass[12pt, twocolumn]{IEEEtran}

\usepackage[utf8]{inputenc}
\usepackage[T1]{fontenc}
\usepackage{amsmath}
\usepackage{siunitx}
\usepackage{graphicx}
\usepackage{chemformula}
\usepackage{soul}
\usepackage{float}
\usepackage{physics}
\usepackage{pifont}
\usepackage{cite}
\usepackage{hyperref}

\newcommand{\U}[1]{\mathrm{#1}}
\DeclareMathOperator\erfc{erfc}
\DeclareMathOperator\SNR{SNR}
\DeclareMathOperator\SER{SER}

\graphicspath{ {./} }

\usepackage{tikz}
\usepackage{textcomp}

\newcommand\copyrighttext{%
  \footnotesize This work has been submitted to the IEEE for possible publication. Copyright may be transferred without notice, after which this version may no longer be accessible.}
\newcommand\copyrightnotice{%
\begin{tikzpicture}[remember picture,overlay]
\node[anchor=south,yshift=10pt] at (current page.south) {\fbox{\parbox{\dimexpr\textwidth-\fboxsep-\fboxrule\relax}{\copyrighttext}}};
\end{tikzpicture}%
}

\begin{document}

\bstctlcite{BSTcontrol}
\title{Fundamental Performance Limits\\on Terahertz Wireless Links Imposed by\\Group Velocity Dispersion}

\author{Karl~Strecker,
        Sabit~Ekin,~\IEEEmembership{Member,~IEEE,}
        and~John~O'Hara,~\IEEEmembership{Senior Member,~IEEE}%
\thanks{K. Strecker, S. Ekin, and J. O'Hara are with the Department
of Electrical and Computer Engineering, Oklahoma State University, Stillwater,
OK, 74078 USA}%
\thanks{Manuscript submitted to arXiv April 26, 2021.}}

\markboth{ }%
{ }

\maketitle
\copyrightnotice
\begin{abstract}
A theoretical framework and numerical simulations quantifying the impact of atmospheric group velocity dispersion on wireless terahertz communication link error rate were developed based upon experimental work.  We present, for the first time, predictions of symbol error rate as a function of link distance, signal bandwidth, signal-to-noise ratio, and atmospheric conditions, revealing that long-distance, broadband terahertz communication systems may be limited by inter-symbol interference stemming from group velocity dispersion, rather than attenuation. In such dispersion limited links, increasing signal strength does not improve the symbol error rate and, consequently, theoretical predictions of symbol error rate based only on signal-to-noise ratio are invalid for the broadband case. This work establishes a new and necessary foundation for link budget analysis in future long-distance terahertz communication systems that accounts for the non-negligible effects of both attenuation and dispersion.
\end{abstract}

\begin{IEEEkeywords}
Millimeter wave communication, Chromatic dispersion, Bit error rate, Atmospheric modeling, Quadrature amplitude modulation, Constellation diagram.
\end{IEEEkeywords}

\IEEEpeerreviewmaketitle

\section{Introduction}
\IEEEPARstart{G}{lobal} data consumption is on the rise\cite{wells_2009, cherry_2004}. The demand for data rate has followed an exponential growth pattern over the last decade, and is projected to continue this trend over the decade to come\cite{mumtaz2017terahertz}. Disruptive technologies such as virtual reality \cite{qi2016quantifying}, telemedicine, and machine-to-machine communication in the Internet of Things\cite{dastjerdi_buyya_2016} are expected to produce and consume data at an unprecedented rate, and there are strong indicators that current communication infrastructure - even emerging 5G communication systems - will be unable to handle this influx of traffic\cite{mumtaz2017terahertz, cisco_2019}.

At the same time, the question of how to rapidly and economically supply rural communities and remote industrial sites with ``last-mile'' broadband internet access has been highlighted by the recent COVID-19 pandemic, which has rapidly accelerated the use of digital technology in ``work from home'' applications\cite{pandey2020impact}. In many cases, trenching a fiber optic connection to an isolated community or remote industrial site is not cost effective, and point-to-point microwave backhaul links do not have the bandwidth required to handle the data rate requirements of the future, where bandwidth-hungry activities such as high-resolution video streaming of lectures and business meetings are anticipated to be widespread\cite{favale2020campus}.

Terahertz wireless communication links may offer a solution to this problem:\cite{song_nagatsuma_2011} the terahertz communication channels (0.1 to 1~THz) offer exceptionally large bandwidths and theoretically allow data to be transmitted wirelessly at rates rivaling that of some fiber optics. Furthermore, setting up a wireless terahertz link is much faster and less costly than trenching fiber optic cable (particularly over difficult terrain), making it attractive for rapid, economical high-bandwidth connectivity.

The greatest drawback of using terahertz links to provide remote, high-bandwidth connectivity is that terahertz radiation is strongly attenuated by both atmospheric oxygen and water vapors. It is sometimes assumed that this will severely limit, or even prohibit, long-distance (multi-kilometer) terahertz wireless communication. However, over the last decade terahertz communication link prototypes have achieved high bandwidth communication at multi-kilometer distances by operating within the broad atmospheric windows in the 0.1 to 0.3 THz range, where atmospheric attenuation is minimized. For example, in 2015, a 10~Gbps link spanning 5.8~km using BPSK (binary phase shift keying) modulation and a carrier frequency of 120~GHz was demonstrated \cite{hirata20105}, as was a 5~Gbps link spanning 21~km using 16-QAM (quadrature amplitude modulation) at 140~GHz in 2017\cite{Wu_2017}. Over shorter distances, higher data rates have been achieved, such as a 64~Gbps link demonstrated in 2015 that spanned 0.85~km with a carrier frequency of 240~GHz\cite{kallfass201564}. As terahertz communication technology continues to advance, increasingly powerful transmitters and more sensitive receivers will no doubt enable communication over longer distances and at higher data rates than what is currently possible. It is reasonable to envision that a 64~Gbps link could one day span 10 or even 20~km, providing not only last-mile connectivity, but also broadband wireless backhaul service.

While limitations arising from bulk atmospheric attenuation may be overcome through increasingly powerful transmitters, the atmosphere also reshapes broadband waveforms through both spectral absorption \cite{jornet2011channel, han2014multi, han2016distance} and group velocity dispersion (GVD)\cite{Mandehgar_2014, Mandehgar_2015, strecker2020compensating}. Spectral absorption and GVD are distinctly different mechanisms of temporal dispersion, but within the broad atmospheric windows discussed here, spectral absorption is minimized and GVD is the primary concern. As such, GVD is the focus of this work. GVD causes a temporal broadening of transmitted data symbols by changing the propagation velocity of electromagnetic radiation unequally with frequency, and sufficiently severe GVD may cause neighboring communication symbols to merge, making them irresolvable in an undesirable phenomena known as intersymbol interference (ISI). GVD-induced ISI becomes more severe in proportion to both signal bandwidth and transmission distance – meaning that future terahertz links with high bandwidth-distance products will experience significant GVD-induced bit errors, very much unlike extant wireless communication technologies. Although the broadening of high-bandwidth terahertz pulses by GVD has been precisely modeled and experimentally verified \cite{yang2011broadband}, GVD in terahertz \emph{communication links} is largely neglected in published terahertz literature, with a few exceptions\cite{Mandehgar_2014, strecker2020compensating}.

One such exception is the work of Mandehgar \textit{~et al.} in 2014, which showed in some cases that GVD – not absorption – becomes the limiting factor in the maximum bandwidth-distance product of a terahertz link\cite{Mandehgar_2014}. Their work makes predictions, based on graphical analysis of impulse broadening, about data rate limits for various terahertz channels operating with an impulse radio modulation scheme, but no calculations are provided for more spectrally efficient modulations such as quadrature amplitude modulation (QAM). More seriously, as Mandehgar \textit{et al.} themselves point out, their work does not account for the inter-related effects of signal-to-noise ratio ($\SNR$) or maximum allowable symbol error rate ($\SER$) and achievable bitrate.

While non-statistical methods based on impulse broadening demonstrate the physics of dispersion, the lack of a theoretical framework for analyzing GVD's impact on SER that incorporates the effects of noise, and the lack of statistical simulations using modulated data streams transmitted over a dispersive channel to validate the framework, represent critical knowledge gaps to be filled\cite{ma_shrestha_moeller_mittleman_2018}.

This work fills that knowledge gap by providing a theoretical framework, coupled with experimentally-founded and consistent simulations, for predicting the SER of a terahertz link in the presence of both noise and dispersion. It shows how GVD produces \emph{deterministic} limits on SER as a function of link propagation distance, atmospheric conditions, and signal bandwidth, but also how \emph{stochastic} noise interacts with dispersion in unexpected ways to dramatically alter link performance. This work leverages high-fidelity models of the atmosphere to produce numerous case studies of the transmission of modulated data over the terahertz channel centered at 250~GHz. The results of these simulations allow us to draw fundamental and highly practical and quantitative conclusions about the impact of atmospheric GVD on broadband wireless terahertz data links.

The remainder of the work will be structured as follows: first, we will describe the methodology used to model the atmospheric channel and how a theoretical prediction of symbol error rate can be derived from the channel model. Next, we will present a comparison of the theoretical error rate predicted by the model to that observed in numerical simulation. Finally, we will discuss the behavior predicted by the dispersion model, and conclude with the implications for long-distance, high-bandwidth terahertz wireless links.

\section*{Methodology}
\subsection{Modeling the Channel Transfer Function}
A theoretical treatment of the effects of GVD begins with the transfer function of the channel. Molecular response theory (MRT) \cite{yang2014determination}, in conjunction with the HITRAN spectroscopic database \cite{gordon_2017}, was used to model the atmospheric channel from 0.1 to 1~THz as a function of temperature, pressure, and atmospheric gas vapor densities. The frequency, resonance strength, and broadening factor of each molecular resonance for both water vapor and oxygen was extracted from the database, selecting only those resonances near enough and strong enough (including all resonances up to approximately 5~THz) to influence channels between 0-1 THz. For each resonance, the extracted parameters of resonance frequency, resonance strength, and broadening factor were used to calculate the collision-broadened line shape in accordance with MRT. The shape of each broadened line depends heavily on the temperature, pressure, and vapor density of the atmosphere, which necessitates recomputing each linewidth whenever one of these parameters is changed. After the linewidth of every resonance has been computed, the individual line shapes are summed over the frequency bands of interest, resulting in the frequency-dependent transfer function of the atmosphere, which takes the general form $H_a (\omega)=\alpha(\omega)\exp[{-j\phi(\omega)}]$, where $\alpha(\omega)$ and $\phi(\omega)$ are the frequency-dependant attenuation and phase shift imparted by the atmosphere, respectively, and $j=\sqrt{-1}$. This process is the same as the methodology of Mandehgar \textit{et al.} in their 2015 work on atmospheric dispersion\cite{Mandehgar_2015}. This modeling procedure fully accounts for the atmosphere’s action on both the amplitude and phase of the propagating terahertz signal, including higher order dispersion, replicating the effects of the atmosphere to a high degree of accuracy\cite{OHara_2018}. Readers are referred to previous work\cite{yang2012understanding} for the mathematical development of atmospheric transfer functions.

In general, both $\alpha(\omega)$ and $\phi(\omega)$ result from the summation of thousands of broadened molecular resonances, and cannot be concisely expressed as analytic formulas. Consequently, when discussing temporal dispersion (a phase effect), it is convenient to expand $\phi(\omega)$ as a Taylor series around the center frequency $\omega_0$. The atmospheric phase shift is thus represented as

\begin{align}
\phi(\omega)
    & = \sum_{n=0}^{\infty} \phi_n \frac{(\omega-\omega_0)^n}{n!}.
\end{align}

In the expansion of phase, $\phi_0$ (units of $rad$) and $\phi_1$ (the linear phase term, with units of $rad\cdot s$) are inconsequential in reshaping (or broadening) the signal. However, $\phi_2$ (the quadratic term, with units of $rad\cdot s^2$) produces a nonlinear phase shift over frequency and will reshape the signal, as will all higher-order nonlinear phase terms ($\phi_3$, $\phi_4$, etc). The $\phi_2$ term is commonly called group delay dispersion (GDD), or second-order dispersion. It is equivalent to GVD, but it is not normalized to propagation distance, and it is the dominant factor in dispersing the signal. This phase shift, the atmospheric attenuation $\alpha(\omega)$, the communications system's filter characteristics, and the injected thermal noise all form the channel over which the communication signal propagates.

\subsection{Modulation-Dependant Impact of the Chanel}
\begin{figure}[ht]
\centering
\includegraphics[ width=0.30\textwidth]{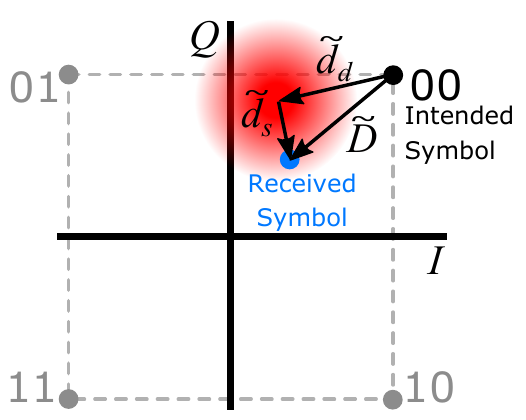}
\caption{Constellation diagram (real versus imaginary plot) of a 4-QAM modulation scheme showing how deterministic dispersion $\tilde{d}_d$, and stochastic noise $\tilde{d}_s$ contribute to displace the received symbol to $\tilde{D}$ from its intended location on the IQ plane. The red shading indicates the qualitative variable range of $\tilde{d}_s$ due to stochastic processes. }
\label{fig:A}
\end{figure}

In a generic $m$-QAM communication system, $m$ is said to be the modulation order, and each of the $m$ possible communication symbols is represented by a unique combination of the magnitude and phase angle of the carrier wave. However, both GVD-induced ISI and thermal noise will alter the magnitude and phase of each transmitted symbol by the time it arrives at the receiver, such that the received amplitude and phase are no longer their ideal values. This can be visualized by means of a constellation diagram. The combined effects of noise and GVD-ISI result in a complex net displacement $\tilde{D}$ of the received symbol from its original (intended) value, which is in part stochastic and in part deterministic (see Fig.~\ref{fig:A}). We use a tilde notation as a reminder that this displacement is a complex quantity, having both phase and amplitude effects. For every symbol, then, the displacement $\tilde{D}$ can be described as $\label{eqn:total_displacement}\tilde{D} =\tilde{d}_d+\tilde{d}_s$, where $\tilde{d}_d$ and $\tilde{d}_s$ are the deterministic and stochastic contributions to displacement, due to dispersion and noise, respectively.

\begin{figure}[ht]
\centering
\includegraphics[ width=0.49\textwidth]{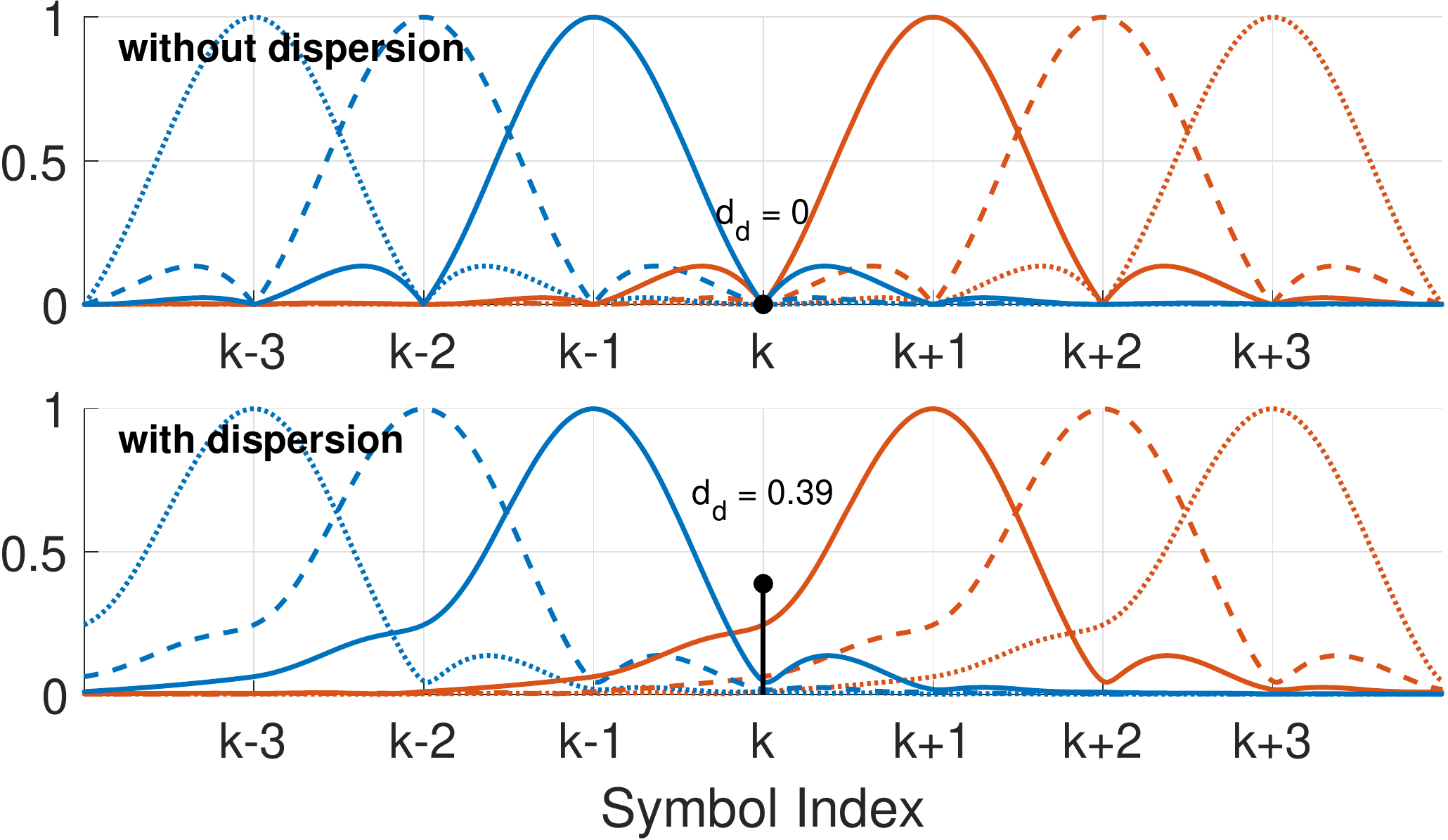}
\caption{\label{fig:B}A sequence of noiseless, normalized channel impulse response envelopes, demonstrating how the sampled value of a symbol $k$ is influenced by neighboring symbols in both dispersionless (top) and dispersed (bottom) cases for a symbol span of $p=3$. Dispersion-broadened pulse wings overlap to produce a dispersion displacement $d_d$ as a function of both preceding (blue) and subsequent (red) symbol value. The symbol envelopes shown are from a 20~GBd link at 250~GHz over 10~km, with a raised cosine filter rolloff of $\beta=0.5$.
}
\end{figure}

As we will now explain, the deterministic component of displacement for any particular symbol directly depends both on the channel properties and on the values of the preceding and subsequent symbols of the data stream, leading to different values of $\tilde{d}_{d}$ for each received symbol. This is illustrated in Fig.~\ref{fig:B}, which shows the noise-free, normalized channel impulse response (magnitude) of identical symbols (three preceding and three following) around the symbol of interest, the $k^\U{th}$ symbol. Note that these have been frequency-shifted to baseband to remove the rapid carrier oscillations that would otherwise obscure the figure. Due to normalization, the shape of each impulse response is mostly governed by the typical raised cosine filtering scheme. Under non-dispersive conditions, the top figure shows how every neighboring symbol provides zero contribution to the $k^\U{th}$ symbol displacement, $\tilde{d}_{dk}$, at the exact sample point, which is, of course, the purpose of raised cosine filtering. However, under dispersive conditions the bottom figure shows how these impulse responses spread beyond their assigned time slots and coherently superpose their dispersed ``wings'' onto the $k^\U{th}$ symbol. The complex sum of all of these wings is the dispersive displacement $\tilde{d}_{dk} $ of the $k^\U{th}$ symbol. Note that the alteration of the impulse response directly and deterministically results from the aforementioned $\phi_2$ and higher-order elements of the atmospheric transfer function. 

\subsection{Computation of the Dispersion Displacement}
To use $\tilde{d}_d$ in a statistical SER study, this phenomenon must be generalized to the channel conditions and all possible symbol combinations. Rigorously, the dispersion displacement $\tilde{d_d}$ of the $k^{\U{th}}$ symbol in the sequence is a complex weighted sum of \emph{all} preceding and subsequent symbols. However, in practice, GVD-ISI is dominated by the influence of only the nearest symbols, while the wings of distant symbols are of such low amplitude as to be dwarfed by thermal noise. The infinite weighted sum may thus be truncated to span only the $p$ symbols near enough to significantly influence the displacement of the $k^\U{th}$ symbol. The truncated summation for the $k^\U{th}$ symbol may thus be expressed as
\begin{equation}
 \label{eqn:dispersion_displacement_sum_truncated}
 \tilde{d}_{dk} = \sum_{n=1}^{p} \tilde{s}_{k+n-1-p}\tilde{w}_{n-1-p} + \tilde{s}_{k+n}\tilde{w}_{n},
\end{equation}
\noindent where $p$ is the integer number of symbols (usually 3 or 4) spanned by the summation in both the $+$ and $-$ directions around $k$. To avoid negative indices and simplify future expressions, this can be re-expressed as the scalar product of two complex vectors
\begin{equation}
 \label{eqn:dispersion_displacement_vector}
 \tilde{d}_{dk} = \mathbf{S} \cdot {\mathbf{W}} =\sum_{n=1}^{2p} \tilde{s}_{n}\tilde{w}_{n}.
\end{equation}
\indent In this notation, the elements are assumed to be centered around the $k^\U{th}$ symbol, $\tilde{w}$ in ${\mathbf{W}}$ are the individual complex symbol weights (corresponding to the symbols $\tilde{s}$ in $\mathbf{S}$), and the indices of the elements in ${\mathbf{W}}$ range from $1$ to $2p$. The indices greater than $p$ indicate weights and symbols following the $k^\U{th}$ symbol, while indices 1 to $p$ denote weights and symbols preceding the $k^\U{th}$. Also note that the $k^{\text{th}}$ (center) symbol itself is not included in the summation, since we are only interested in the relative displacement of the symbol, not its absolute value.

The weighting factors $\tilde{w}$ in ${\mathbf{W}}$ depend only on the characteristic (impulse) response of the channel, which depends on atmospheric conditions and therefore changes very slowly compared to the transmitted symbol rate. As a result, the elements of ${\mathbf{W}}$ may be assumed constant for SER calculations, however we note that they would slowly change over long durations (multiple minutes). The values of the neighboring symbols themselves, however, change with every symbol transmitted, but may only take on a limited number of discrete values. Consequently, for a given set of atmospheric channel conditions, there is a limited number of discrete displacements that GVD-induced ISI can feasibly impart to a transmitted symbol. Specifically, if only the nearest $p$ symbols in each direction influence the displacement, then there will be $m^{2p}$ possible dispersion displacements.

To find all the possible displacements, we must simply repeat the calculation of equation (\ref{eqn:dispersion_displacement_sum_truncated}) for all $m^{2p}$ possible combinations of symbol values. This gives every possible $\mathbf{S}$, and therefore every possible dispersion shift for the channel. The entire operation may be expressed in compact matrix notation as
\begin{equation}
 \mathbf{d}_d = \overline{\mathbf{S}}{\mathbf{W}} = 
\begin{bmatrix}
 \tilde{s}_{1,1} & \tilde{s}_{1,2} & \cdots & \tilde{s}_{1,2p} \\
 \tilde{s}_{2,1} & \tilde{s}_{2,2} & \cdots & \tilde{s}_{2,2p} \\
 \vdots & \vdots & \ddots & \vdots \\
 \tilde{s}_{m^{2p},1} & \tilde{s}_{m^{2p},2} & \cdots & \tilde{s}_{m^{2p},2p} 
 \end{bmatrix}
 \begin{bmatrix}
 \tilde{w}_{1} \\
 \tilde{w}_{2} \\
 \vdots\\
 \tilde{w}_{2p}
 \end{bmatrix},
\end{equation}
where $\mathbf{d}_d$ is the set of all possible $d_d$, $\overline{\mathbf{S}}$ is a $(m^{2p}\cross2p)$ matrix containing every possible combination of the $2p$ neighboring symbols, and $\mathbf{W}$ remains a $(2p\cross1)$ vector containing the complex weights. Each element of $\mathbf{d}_d$ is one of $m^{2p}$ possible complex valued dispersion displacements, so that if a complex-valued symbol $X$ is transmitted, the receiver will receive one of the values $Y = X + \mathbf{d}_d$. As can be seen, $X+\mathbf{d}_d$ will depend on the atmospheric conditions through $\mathbf{W}$ and on the particular symbol sequence surrounding $X$ through $\overline{\mathbf{S}}$. 

\subsection{The Stochastic Contribution of Thermal Noise}
Having determined the set of all possible deterministic displacements due to GVD, it is now necessary to determine the stochastic contribution to displacement, $d_s$ due to thermal noise. Typically, theoretical treatments of SER as a function of SNR assume that additive white Gaussian noise randomly displaces the received symbols from a single, ideal constellation point. However, in the case of a signal propagating through a temporally dispersive medium, the thermal noise does not displace received symbols around a single, ideal symbol $X$, but rather around one of the possible complex-valued, dispersion-shifted locations $X + \mathbf{d}_d$, as illustrated in Fig.~\ref{fig:A} by the red shaded area. This stochastic displacement has a component in $I$ and a component in $Q$, or $\tilde{d}_s = d_{sI}+jd_{sQ}$, which can be treated independently and then combined because they are uncorrelated. The probability density functions governing the I- and Q-components due to white Gaussian noise are expressed as
\begin{align}
 f(I,I_0) = \frac{1}{\sqrt{N_0\pi}}\exp\left[-\frac{(I - \sqrt{E}I_0)^2}{N_0}\right],\\
 \quad g(Q,Q_0) = \frac{1}{\sqrt{N_0\pi}}\exp\left[-\frac{(Q - \sqrt{E}Q_0)^2}{N_0}\right],
\end{align}
\noindent where $N_0$ is the noise power, $E$ is the signal power, $I_0 = \mathrm{Re}(X+d_d)$, and $Q_0 = \mathrm{Im}(X+d_d)$. At this point, a specific modulation scheme must be assumed in order to proceed with the derivation. Now and throughout the rest of this work, we assume 4-QAM modulation is used. The probability that a 4-QAM modulated symbol will be received correctly is the probability that the symbol will fall into any one particular quadrant of the complex plane. Without loss of generality, we can choose the quadrant covered by angles from $\theta = 0$ to $\theta = \frac{\pi}{2}$ as the set of all points $C$ within which the received symbol must fall in order to be correctly decoded. Thus, with the ideal symbol value is $X = \mathrm{e}^{j\frac{\pi}{4}}$, the probability $P$ that the received symbol $Y$ falls outside the correct decision region for a given $d_d$ is
\begin{align}
 & P(Y \notin C | X,d_d) = \nonumber\\
    & 1 - \left[1-\int_{-\infty}^{0} f\left(I,\mathrm{Re}(X+d_d)\right)\;\mathrm{d}I\right]\nonumber\\  
    & \times \left[1-\int_{-\infty}^{0} g\left(Q,\mathrm{Im}(X+d_d)\right)\;\mathrm{d}Q\right],
\end{align}
\noindent or, with the integrals expressed in terms of the complimentary error function $\erfc(z)$,
\begin{align}
\label{probability_equation_error}
 P(Y \notin C | X,d_d) & = \frac{1}{4}\erfc\left[\sqrt{\frac{E}{N_0}}\left(\Re(X+d_d)\right)\right]\nonumber\\
 & \times \erfc\left[\sqrt{\frac{E}{N_0}}\left(\Im(X+d_d)\right)\right].
\end{align}

Finally, if we assume the data sent over the communication channel is random, then all symbol combinations – and thus all potential dispersion shifts $d_d$ in $\mathbf{d_{d}}$ – are equally likely. The total probability of error is then the arithmetic mean of equation~(\ref{probability_equation_error}) evaluated at every one of the $m^{2p}$ dispersion shifts $d_d$ in $\mathbf{d}_d$, or
\begin{equation}
    \SER = \frac{1}{m^{2p}}\sum_{n=1}^{m^{2p}}P(Y \notin C | X,d_{d_n}).  
    \label{modeled_SER_total}
\end{equation}
 \indent Importantly, equation~(\ref{modeled_SER_total}) gives a theoretical prediction of SER as a function of SNR and incorporates the potentially significant effects of GVD, in contrast to previous models. Because of the number of calculations involved and the need to compute the value of the non-elementary complementary error function, this theoretical prediction is best evaluated numerically.

\begin{figure}[ht]
\centering
\includegraphics[ width=0.48\textwidth]{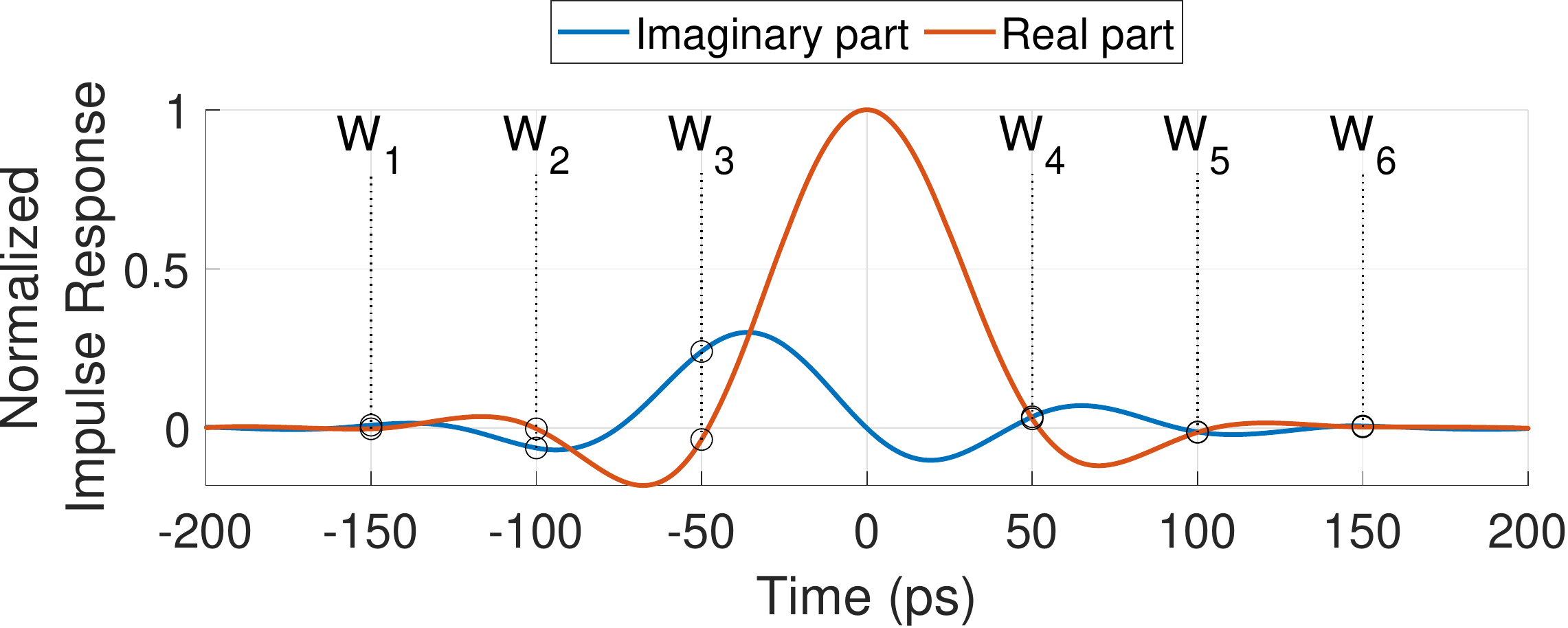}
\caption{Normalized baseband channel impulse response of a 20~GBd signal, sampled at intervals of $T_s$ to obtain the elements of $\mathbf{W}$. The impulse response includes the effects of both the atmosphere and the raised cosine pulse-shaping filter, which here has a rolloff factor of $\beta=0.5$.}
\label{fig:FIG_C}
\end{figure}

\subsection{Calculating the Elements of \textbf{W}}
The method described above depends on accurate values for the elements of $\mathbf{W}$. The elements of $\mathbf{W}$ encapsulate the deterministic action of the communication channel on the transmitted symbols, and so must account for the present atmospheric conditions, the transmit and receive filter characteristics, and any other deterministic factors which affect the signal, intentionally or unintentionally. The elements of $\mathbf{W}$ are obtained by evaluating the normalized baseband impulse response of the channel $h(t)$ at integer multiples of the symbol period, $T_s$. These sample times correspond to the time-domain locations of the preceding and subsequent symbols; Fig. \ref{fig:FIG_C} illustrates how the elements of $\mathbf{W}$ are found from $h(t)$. First, $h(t)$ is determined by either direct measurement or by inverse Fourier transform of the channel transfer function $H(\omega)$. $H(\omega)$ may also be found by direct measurement or by using known channel filter characteristics and numerical resonance calculations from spectroscopic databases (as is the case in this work). The impulse response is then frequency shifted to baseband and broken into real and imaginary parts, which are finally sampled to yield the complex weights. 

\section*{Results}
\begin{figure}[!t]
\centering
\includegraphics[width=0.49\textwidth]{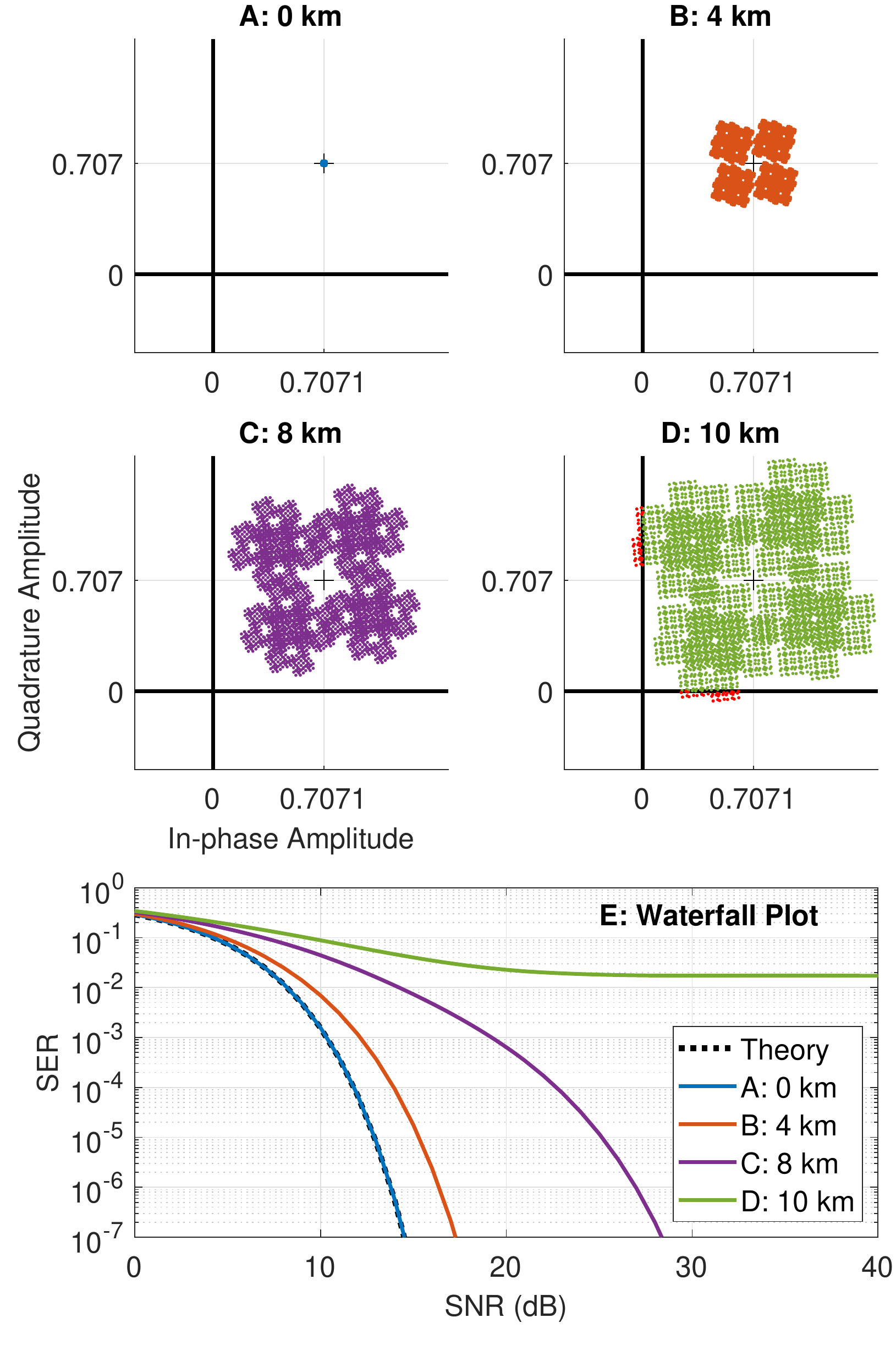}
\caption{Constellation diagrams of all elements $d_{d}$ in $\mathbf{d}_d$, and a corresponding waterfall plot of the predicted error performance, for link distances of 0, 4, 8, and 10 km, a symbol span $p=3$ and a filter rolloff factor $\beta=0.5$. Constellation diagrams show an intended symbol value of $X=\mathrm{e}^{j\frac{\pi}{4}}$, with infinite SNR. Plots are for a 30 GBd link centered at 250 GHz with water vapor density $\rho_{\rm wv}~=~10.37$~g/m$^3$ (60\% relative humidity at \SI{20}{\degreeCelsius}). For the 10 km link (case D), temporal dispersion is so severe that errors are deterministically produced - these are indicated by the red constellation points on constellation diagram D.}
\label{fig:D}
\end{figure}

By evaluating equation~(\ref{probability_equation_error}) for all dispersion shifts in $\mathbf{d}_d$, we produced predictions of SER as a function of SNR for a 250~GHz center frequency terahertz wireless link over various distances and bandwidths. The 250~GHz carrier is attractive for terahertz wireless communication due to its location in the center of a very broad atmospheric window, where both attenuation by water vapor and corresponding GVD are at a relative minimum and available bandwidth is in excess of 100~GHz. A prediction of SER versus SNR for a 30 gigabaud (GBd) link is shown in Fig.~\ref{fig:D}, which shows all possible received symbols $Y=X+\mathbf{d_d}$ (absent of noise) in the constellation diagrams, and also shows ``waterfall plots'' of SER vs SNR (inclusive of noise) for four link configurations (cases A through D). In the figure, the 0~km link (case A) suffers no dispersion, and consequently the SER vs. SNR curve matches the well-known expression for a narrowband (dispersionless) 4-QAM link \cite{goldsmith2005wireless}, namely,
\begin{equation}
 \SER_c=\erf{\sqrt{\frac{\SNR}{2}}} - \frac{1}{4}\left(\erf{\sqrt{\frac{\SNR}{2}}}\right)^2,
 \label{eqn:theoretical_SER_4-QAM}
\end{equation}
where $\SNR$ is the signal-to-noise ratio of the signal. In addition, the received symbols match the transmitted symbols ($Y=X$) in the noise-free constellation diagram.

However, for the case where the link distance is significant (cases B and C), dispersion displacement increases the probability of symbol errors by shifting many of the received symbols closer to the decision boundaries ($I=0, Q=0$). This ``blooming'' of the received symbols due to dispersion is evident in the constellation diagrams. This also causes the SER vs SNR curve to deviate from that predicted by the classical (dispersionless) model, $\SER_c$, such that increased signal strength is required to achieve the same SER. Significantly, in the case where the path length is sufficiently long (case D) dispersion, by itself, displaces some symbols over the decision boundaries. If this is the case, those symbols will \emph{always} be received in error at high SNR values, setting a ``dispersion limit'' beyond which the SER cannot be improved by increasing signal strength.

Empirical validation of the behavior predicted by the waterfall plot of Fig.~\ref{fig:D} was not feasible due to the difficulties associated with constructing a 10~km long climate-controlled terahertz channel for 30~GBd communication. Accordingly, the predictions made by equation~\ref{probability_equation_error} were corroborated by numerical simulation. The simulations was performed by mathematically modeling all major system components (data generation, 4-QAM modulation, pulse shaping, carrier modulation, propagation through the atmospheric channel, homodyne detection, receive filtering, and demodulation), which were used to simulate the transmission and reception of a pseudo-random data sequence comprised of 4000-bit frames. At the receiver, errors were identified by comparing the transmitted and received versions of each data frame. By repeating this process for 2000 frames, a statistical SER for the link was predicted. Most of the details regarding how the simulation was performed are of little relevance to this work - the definition of a 4-QAM communication system in software is well-known. However, a few clarifying details regarding signal attenuation need to be discussed.

The channel model used in the simulation must account for two main attenuation effects experienced by the signal, those being the frequency-dependent change in amplitude caused by molecular vapor resonances, and the net decrease in SNR due to other factors. As in previous work \cite{yang2012understanding}, the frequency-dependent change in amplitude was modeled by multiplication of the transmitted signal’s complex amplitude spectrum, $E_\U{in}(\omega)$, with the complex transfer function of the atmosphere, $H_a (\omega)$, which was previously described. The second of the two channel effects - the net decrease in $\SNR$ - is caused by a collection of factors that may include the transmitter power, receiver sensitivity, antenna directivity, space loss, and filter characteristics\cite{Agrawal_2010}. Rather than attempt to account for all such specifics, we set the $\SNR$ of the signal directly by injecting a prescribed amount of additive white Gaussian noise into the system immediately prior to demodulation at the receiver. This overall approach abstracts away the details of the link hardware, making the simulation results applicable to arbitrary terahertz communication systems while still accounting for the frequency-dependent effects of the atmosphere.

\begin{figure}[ht]
\centering
\includegraphics[width=0.49\textwidth]{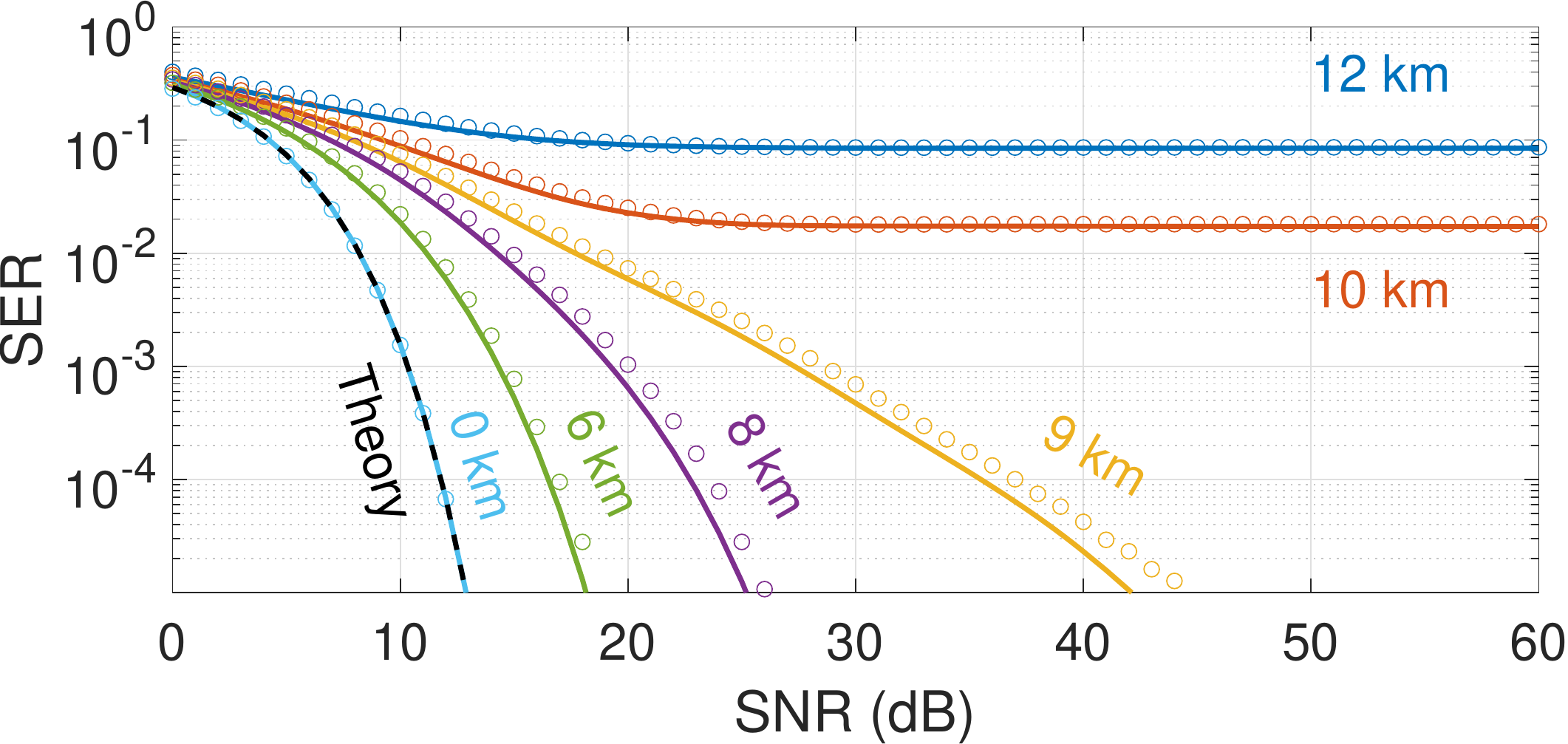}
\caption{Comparison of error rates predicted by the model to those observed in simulation for a 30~GBd terahertz link centered at 250~GHz, transmitting through the atmosphere with water vapor density $\rho_{\rm wv}~=~10.37$~g/m$^3$ (60\% relative humidity at \SI{20}{\degreeCelsius}), at path lengths of 0, 6, 8, 9, 10, and 12~km using a filter rolloff $\beta=0.5$. Solid lines show the error rates predicted by the model (using a symbol span $p=3$), and the open circles show the error rates computed by simulation. The theoretical SER curve for a classical (dispersionless) 4-QAM signal (equation (\ref{eqn:theoretical_SER_4-QAM})) is also shown for reference (dotted line labeled ``Theory'').}
\label{fig:F}
\end{figure}

In Fig.~\ref{fig:F}, the simulation results are compared to those predicted by the model. The figure shows very good qualitative and quantitative agreement between the error rates predicted by the model (solid lines) and those observed in numerical, stochastic simulations (open circles). There is slight disagreement between the model and simulation toward higher SNR values for the 8~km and 9~km links, where the simulation observes a higher number of errors than predicted by the model. This initially seems due to the truncated symbol span of $p=3$, which slightly underpredicts the maximum possible dispersion displacement. Nevertheless, the general agreement gives us confidence that the model properly describes the underlying mechanisms involved and accurately predicts the error rate performance of broadband terahertz links affected by GVD.

\section*{Discussion}

From the plots in Fig.~\ref{fig:F} it is clear that $\SER$ is affected by both dispersion and attenuation, and in general, neither is unequivocally dominant: \emph{either} may be the principal channel mechanism that determines the lower bound on $\SER$. Case studies based on Fig.~\ref{fig:F} provide important details surrounding this fact. In these discussions, signal ``bandwidth'' is interchangeable with symbol rate, since $x$~GBd corresponds to $x$~GHz of bandwidth in 4-QAM modulation (which has a spectral efficiency of 2~bits/s/Hz).

Fig.~\ref{fig:F} shows $\SER$ plotted explicitly against $\SNR$ for a 30~GBd link. For the dispersionless case of a narrowband signal, the exact bandwidth is irrelevant so long as the signal lies within the semi-constant absorption window in the atmosphere, where reduction in $\SNR$ due to signal attenuation is the primary effect on the signal. In Fig.~\ref{fig:F}, the dotted line labeled ``Theory'' (equation (\ref{eqn:theoretical_SER_4-QAM})) corresponds to this case, and represents the limit placed on $\SER$ by atmospheric attenuation alone.

For a very short path (0~km), dispersion is absent and the $\SER$ predicted by our model matches the established theoretical prediction. However, as dispersion is increased by extending the link distance, our model predicts a deviation from the theoretical relationship between $\SER$ and $\SNR$, where multiple orders of magnitude more errors may occur for a given SNR. Notably, error rate may still be improved by increasing $\SNR$ for link distances of $\le 9$~km.

At or above 10~km however, a significant and fundamental change in behavior is observed in the waterfall curves of Fig.~\ref{fig:F}: $\SER$ becomes decoupled from $\SNR$ such that increasing the $\SNR$ of the signal yields no improvement in the $\SER$, a sharp departure from the behavior predicted by equation (\ref{eqn:theoretical_SER_4-QAM}). This change, at just under 10~km, occurs because dispersion displacement becomes so severe that some some symbols are always received in error. This is illustrated by the red constellation points in Fig.~ \ref{fig:D}, subplot D (10~km constellation diagram). Beyond this point, the physical limitation on link performance is no longer determined by $\SNR$ but by dispersion. Before the dispersion limit, improvements in SER could be gained by increasing the $\SNR$, for example through the use of more powerful transmitters. However, beyond the dispersion limit, improvements in SER require the design and implementation of dispersion-compensating strategies\cite{strecker2020compensating}.

\begin{figure*}[!t]
\centering
\includegraphics[ width=0.9\textwidth]{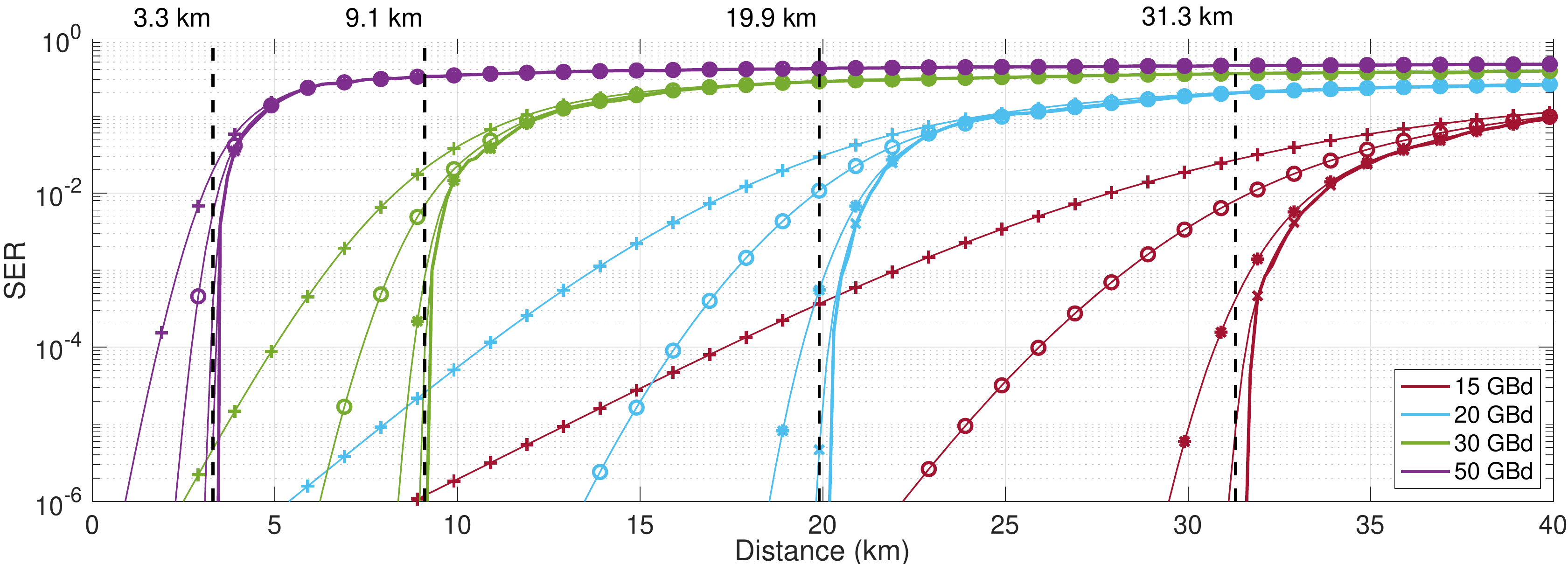}
\caption{Symbol error rate versus distance for a terahertz link centered at 250~GHz, operating at various symbol rates. Atmospheric conditions are water vapor density $\rho_{\rm wv}~=~10.37$~g/m$^3$ (60\% relative humidity at \SI{20}{\degreeCelsius}). The thick solid lines are the noiseless case, and SNRs of 40, 30, 20, and 15 dB are designated by markers \ding{53}, \ding{107}, o, and + respectively. Dispersion limits are indicated by dashed vertical lines.}
\label{fig:RevWaterfallNoisy}
\end{figure*}

The precise value of the dispersion limit may be found by setting the $\SNR$ at the receiver to be infinite (injecting no noise). This corresponds to the physical case of an arbitrarily powerful transmitter, capable overcoming any signal losses. In this condition, the relationship between link distance and SER for various symbol rates is shown by the thick solid lines in Fig.~\ref{fig:RevWaterfallNoisy}.

The results in Fig.~\ref{fig:RevWaterfallNoisy} clearly reveal the dispersion limit on the distance a terahertz wireless link may span, which is shown by the sudden change from error-free operation to $\SER$ values exceeding 1 in 10 for the noiseless case. In some cases, the $\SER$ increases by multiple orders of magnitude over a distance increase of less than 1~km. These dispersion limits are marked on the graph by vertical dashed lines. The critical implication is that no matter how much power a transmitter can emit, nor how sensitive a receiver is developed, there exists a finite distance a terahertz signal of a given bandwidth and carrier frequency can propagate before ISI caused by uncompensated GVD renders the link effectively inoperable. For example, as seen in Fig.~\ref{fig:RevWaterfallNoisy}, a 20~GBd (40~Gbps) link at 250~GHz cannot extend farther than approximately 20~km while maintaining an $\SER$ below $10^{-6}$, while a 30~GBd (60~Gbps) link is restricted to less than approximately 9~km for the same $\SER$, assuming the atmospheric conditions described in the figure caption.

It is also evident from Fig.~\ref{fig:RevWaterfallNoisy} that reductions in bandwidth greatly increase the dispersion limit. In fact, a 10~GBd (20~Gbps) link has a dispersion limit beyond 50~km, and is not visible in Fig.~\ref{fig:RevWaterfallNoisy} for the given set of channel parameters. Such a dispersion limit could be neglected for single-section line-of-sight links because Earth curvature will likely limit the propagation distance. On the other hand, for very high data rates (such as 50 GBd), the dispersion limit is already debilitating at just over 3~km. 

The noiseless scenario is the best-case, when atmospheric attenuation is negligible and thus $\SNR$ is extremely high. Decreasing the $\SNR$ at the receiver due to attenuation has a further detrimental effect on the SER, relaxing the sharp dispersion limits exhibited in the noiseless case, and reducing the distance at which dispersion causes a terahertz link to begin exhibiting symbol errors. Importantly, this means that heavily attenuated signals (that is, signals with low $\SNR$) will suffer an increase in $\SER$ due to \emph{dispersion}, even well below the noise-free dispersion limit. This demonstrates a double penalty imposed by atmospheric absorption in the terahertz bands. Symbol errors are not only increased by reduced SNR, but also by the interplay of noise with dispersion-modified waveforms.

As an illustration of this, refer to the $\SER$ vs. distance plot for the 15~GBd (30~Gbps) link in Fig.~\ref{fig:RevWaterfallNoisy}. When the $\SNR$ is high (40 dB), a 30~Gbps link spanning 25~km is well below the dispersion limit and enjoys error-free operation. However, a reduction in $\SNR$ to 20~dB causes an increase in dispersion-induced errors, and the $\SER$ at 25~km is increased to $3 \times 10^{-5}$. Further decreasing the $\SNR$ to 15~dB raises the error rate at 25~km to $2 \times 10^{-3}$, and pushes the distance for an $\SER$ of $3 \times 10^{-5}$ back to only 15~km.

Fig.~\ref{fig:RevWaterfallNoisy} demonstrates that even at distances well below the dispersion limit, GVD may still significantly impact the performance of a terahertz link if the $\SNR$ is low. It is especially noteworthy that this effect is most pronounced in lower-bandwidth links. From observing only the noiseless plots in Fig.~\ref{fig:RevWaterfallNoisy}, it might be incorrectly assumed that low bandwidth links will not suffer dispersion-induced ISI because they only become dispersion-limited at very long ranges. However, the noisy plots in Fig.~\ref{fig:RevWaterfallNoisy} show that this is not the case: for example, the 15~GBd link shows a dramatic increase in errors due to dispersion when the $\SNR$ is poor, even at distances well below the dispersion limit. Remarkably, a high dispersion limit does not necessarily imply an absence of dispersion-induced symbol errors; it simply means that for most distances, improving the $\SNR$ of the signal will correspondingly improve the $\SER$ of the link.

\section{Conclusion \& Future Direction}

By formulating a theoretical model and performing end-to-end simulations of terahertz digital communication with all atmospheric effects incorporated, it was found that GVD is non-negligible in broadband terahertz links, and GVD limits both transmission distance and $\SER$ in conjunction with atmospheric attenuation. At lower bandwidths (below about 20~Gbps), absorption was found to be the dominant mechanism limiting link distance, which matches the prevailing assumption in the literature. On the other hand, for higher bandwidths, the work reveals that dispersion, and not atmospheric absorption, is the factor that ultimately limits the maximum distance of the link. This work agrees with the observations of Mandehgar et al \cite{Mandehgar_2014}, but also provides a firm analytical foundation for such conclusions and far greater insight in the relationships among $\SER$, bit rate, link distance, and $\SNR$ for high-bandwidth, long distance terahertz links than any other study published to date. 

A significant finding of this work is that theoretical predictions of $\SER$ based only on $\SNR$ are not valid for high bitrate-distance terahertz links. In the literature, it is generally assumed that increasing the power of the terahertz transmitter corresponds to increasing the distance over which data may be reliably transmitted. This may be true in some cases, but only up to a certain point. Once the dispersion limit is reached, GVD becomes the dominant factor is determining link performance, and increasing the $\SNR$ no longer yields any improvement in $\SER$. This means that no matter how powerful a transmitter is developed, the maximum transmission distance of a terahertz link cannot exceed a certain, finite value. Moreover, this dispersion limit on maximum distance decreases rapidly with both increasing bandwidth and increasing vapor density in the atmosphere.

An important, related finding was that the presence of noise exacerbates the negative effects of dispersion at distances \emph{below} the dispersion limit, and this effect is most noticeable in links having a high dispersion limit (\emph{lower} bandwidth). The implication is that dispersion cannot be neglected even below the dispersion limit. Operating below the dispersion limit does not guarantee freedom from GVD-induced ISI, but only that increasing the $\SNR$ of the signal will improve the $\SER$ of the link. 

By revealing that uncompensated dispersion places a physical limitation on the distance of high-bandwidth terahertz communication links, and by demonstrating that may affect link performance well below the dispersion limit, this work has established GVD as a factor that must be included in link budgets for the design and analysis of future broadband terahertz communication links.

\section*{Acknowledgment}
This material is based upon work supported by the National Science Foundation Graduate Research Fellowship Program under Grant No. 1746055. Any opinions, findings, and conclusions or recommendations expressed in this material are those of the author(s) and do not necessarily reflect the views of the National Science Foundation.

\bibliographystyle{./IEEEtran}
\bibliography{./IEEEabrv,./main}

\begin{IEEEbiography}[{\includegraphics[width=1in,height=1.25in,clip]{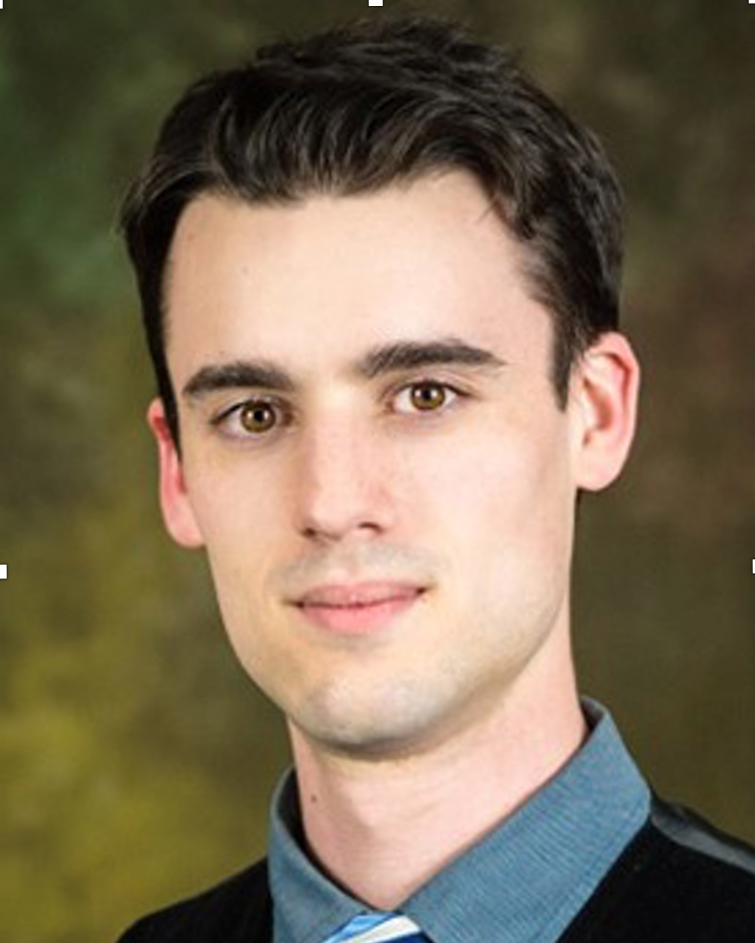}}]{Karl Strecker}
Karl L. Strecker received his B.S. in 2018 and M.S. in 2020, both in electrical engineering from Oklahoma State University.  He is currently pursuing his Ph.D. degree in electrical engineering at the same institution. From 2018 to the present, he has worked as a Research Assistant in the Ultrafast Terahertz and Optoelectronics Laboratory at Oklahoma State University. His research interests include wireless communication, group velocity dispersion management, and terahertz material characterization. Mr. Strecker was a recipient of the 2020 National Science Foundation Graduate Research Fellowship.
\end{IEEEbiography}

\vfill

\begin{IEEEbiography}[{\includegraphics[width=1in,height=1.25in,clip]{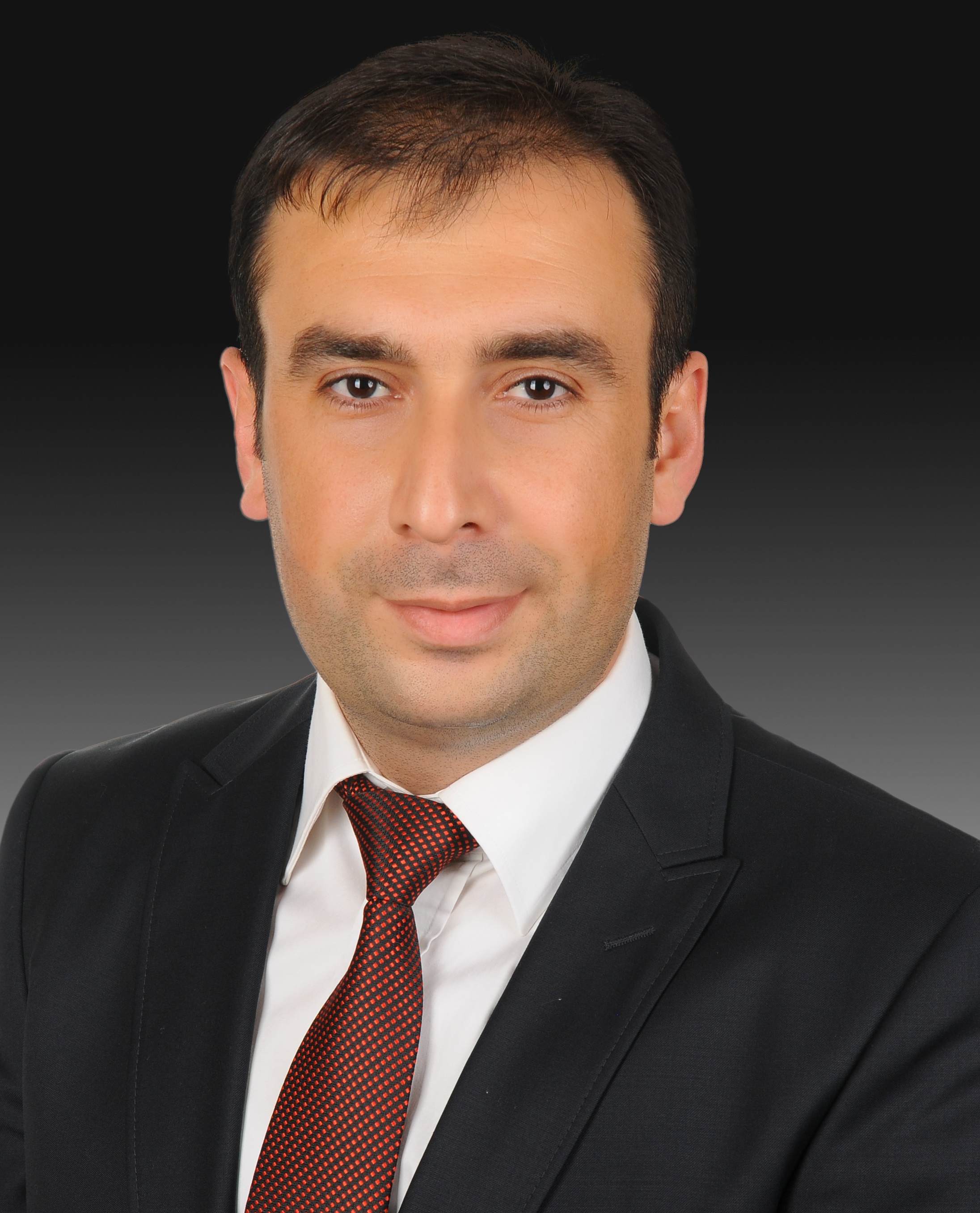}}]{Sabit Ekin}  received the B.Sc. degree in electrical and electronics engineering from Eski\c sehir Osmangazi University, Turkey, in 2006, the M.Sc. degree in electrical engineering from New Mexico Tech, Socorro, NM, USA, in 2008, and the Ph.D. degree in electrical and computer engineering from Texas A\&M University, College Station, TX, USA, in 2012. He was a Visiting Research Assistant with the Electrical and Computer Engineering Program, Texas A\&M University at Qatar from 2008 to 2009. In summer 2012, he was with the Femtocell Interference Management Team in the Corporate Research and Development, New Jersey Research Center, Qualcomm Inc. He joined the School of Electrical and Computer Engineering, Oklahoma State University, Stillwater, OK, USA, as an Assistant Professor, in 2016. He has four years of industrial experience from Qualcomm Inc., as a Senior Modem Systems Engineer with the Department of Qualcomm Mobile Computing. At Qualcomm Inc., he has received numerous Qualstar awards for his achievements/contributions on cellular modem receiver design. His research interests include the design and performance analysis of wireless communications systems in both theoretical and practical point of views, interference modeling, management and optimization in 5G, mmWave, HetNets, cognitive radio systems and applications, satellite communications, visible light sensing, communications and applications, RF channel modeling, non-contact health monitoring, and Internet of Things applications.
\end{IEEEbiography}

\newpage
\vfill

\begin{IEEEbiography}[{\includegraphics[width=1in,height=1.25in,clip,keepaspectratio]{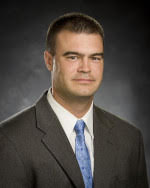}}]{John F. O'Hara} (M’05 - SM’19) received his BSEE degree from the University of Michigan in 1998 and his Ph.D. (electrical engineering) from Oklahoma State University (OSU) in 2003.  He was a Director of Central Intelligence Postdoctoral Fellow at Los Alamos National Laboratory (LANL) until 2006.  From 2006-2011 he was with the Center for Integrated Nanotechnologies (LANL) and worked on numerous metamaterial projects involving dynamic control over chirality, resonance frequency, polarization, and modulation of terahertz waves.  In 2011, he founded a consulting/research company, Wavetech, LLC specializing in automation and IoT devices.  In 2017 he joined OSU as an assistant professor in Electrical \& Computer Engineering.  His current research involves terahertz wireless communications, terahertz sensing and imaging with metamaterials, IoT, and light-based sensing and communications.  He has 3 patents and around 100 publications in journals and conference proceedings.  Dr. O’Hara is a Senior Member of IEEE.
\end{IEEEbiography}

\end{document}